\documentclass{article}%
\usepackage{amssymb}
\usepackage{amsmath}
\usepackage{amsfonts}
\usepackage{graphicx}%
\providecommand{\U}[1]{\protect\rule{.1in}{.1in}}
\begin{document}

\title{Space-time propagator and exact solution for wave equation in a\\layered system}
\author{Victor F. Los* and Nicholas V. Los**
\and *Institute for Magnetism, Nat. Acad. Sci. and Min. Edu. Sci.
\and of Ukraine, Vernadsky Blvd. 36-b, 03142, Kiev, Ukraine
\and **Independent Researcher}

\begin{abstract}
Exact space-time propagator for the wave (second-order in

time) equation in a layered system, made up of a layer sandwiched

between two other different semi-infinite layers, is obtained by

means of the\ multiple scattering theory (MST) approach. The wave

equation for this layered system is characterized by the spatial

dependence of the wave phase velocity which changes in the

perpendicular-to-interfaces direction when crossing the interfaces

and thus having different constant values in different layers of the

threelayer. The MST approach is made possible due to obtained

effective potentials localized at interfaces and responsible

for reflection from and propagation through them. The

obtained space-time propagator exactly solves the considered wave

equation for any initial value and allows for following in time the

processes of reflection and transmission. The solution for an initial

Gaussian wave packet is obtained, analyzed and numerically

visualized for a simple case of the perpendicular-to-interfaces

incoming wave. In particular, the contribution of the backward-moving

components of the wave packet is revealed.

\end{abstract}
\maketitle

\section{Introduction}

Propagation of particles/waves through heterogenous media is a key problem to
be considered in many fields of classical and quantum physics. The
particles/waves' (partial) reflection from and propagation through any
inhomogeneities are well explained by the Huygens-Fresnel principle (first
suggested for the light waves) which describes the wave front propagation in
terms of the wavelets generated by each point of the wave front and their
interference. Multiple scattering of these wavelets at any obstacles results
in the observed phenomena of reflection, refraction and diffraction of waves
in inhomogeneous and disordered media. Thus, the Huygens-Fresnel principle for
waves propagation may be naturally mathematically supported in terms of the
multiple scattering theory (MST). However, this can effectively be done when a
perturbation in the differential equation governing a particle/wave
propagation is localized enough. Particularly, this is the case for electrons
scattered by impurities in disordered systems. A review of MST applications to
solid state-related problems can be found, e.g., in \cite{Gonis and Butler}.

When treating the propagation of waves in heterogenous media, one should often
deal with a potential energy (or refraction coefficient) acquiring different
values in the various macroscopic areas of the system under consideration.
Then, it is generally difficult to consider the potential energy $V(x)$
(refraction coefficient $n(x)$) in different areas as a perturbation because
it is not localized. In such cases the problem is conventionally treated by
imposing certain boundary conditions on the waves (e.g. the continuity of the
electron wave function and its derivative at the interfaces between areas with
different electron potential energy, the boundary conditions for the vectors
of electrical strength and magnetic induction at the interfaces between
different dielectrics, etc). Such conditions may be viewed as mathematical and
serve as a substitute for the MST approach which is conventionally formulated
in terms of a "free" particle/wave Green function (resolvent) modified by the
scatterings at the inhomogeneities such as the interfaces between different
media. The matching approach, however, can be somewhat ineffective in the case
of the wave packet (not the plane wave) propagation which is necessary for
considering the processes of reflection from and transmission through, e.g., a
layered system, in time. Moreover, from the physical point of view it seems
desirable to be able to also apply the multiple scattering theory to this
principally important situation. If the potential energy (refraction
coefficient) changes abruptly (as compared with the wave length), one can
assume that reflection occurs at the single plane (interface) dividing
different materials. Thus, in this case we expect that reflection/transmission
may be described in terms of the MST with a localized at an interface between
areas with different values of potential energy (refraction index)
perturbation potential, which should be found. If so, the MST has an
additional advantage: by adding to this interface effective potential the
potential describing an interface disorder, one will capable of considering
(using the MST framework) the amplitudes of reflection from and transmission
through the intermixed real interfaces which are the key elements for
describing the phenomena in real nanostructures. In such a case, the multiple
scattering by the potential steps and by an interface disorder can be treated
within MST on equal footing. It should also be noted that in multilayers,
multiple scatterings from more than one interface are to be accounted for.
This circumstance makes application of the matching procedure at the
interfaces problematic, especially for the case of real disordered interfaces,
when, except specular scattering, the diffuse scattering should also be taken
into consideration.

Thus, one faces an interesting problem of finding a localized potential
responsible for a particle/wave scattering from a potential inhomogeneity. On
the other hand, from the practical point of view, this rather academic problem
of scattering of a particle/wave from the potential steps has acquired reality
and important practical implementation in the newly emerged field of
nanoscience and nanotechnology dealing, particularly, with multilayer systems.
For example, for electrons, such a situation is realized in layered
nanostructures made of alternating magnetic and nonmagnetic layers, each of a
few atomic layers thick, which has been attracting a good deal of interest
especially since the discovery of the giant magnetoresistance (GMR)
\cite{Baibich et al (1988)} and tunneling magnetoresistance (TMR)
\cite{Jullier (1975), LeClair et al (1994)} effects. Another possible
important application may be the Casimir force \cite{Casimir} emerging
between, e.g., two dielectric plates and caused by the multiple
electromagnetic waves scattering and their interference in the space between
the plates. This effect, principally important from the point of view of
quantum electrodynamics, is often formulated in terms of the MST with the
$\delta$-like scattering potentials localized at the interfaces between a
vacuum and reflecting bodies (see, e.g., \cite{MST Casimir}). Again, one needs
to find these localized potentials in terms of characteristics of the
interface-devided media.

In this paper, we consider a wave propagation through a nonuniform (layered)
media described by the second-order in time differential equation. We focus on
the wave equation, because the case of the first-order in time
(Schr\"{o}dinger) equation was considered earlier. Thus, we will apply the
approach first suggested in \cite{Los 2010} and which allowed for treating the
electron reflection from and transmission through a perfect interface as a
multiple scattering by the interface-localized potentials in the case when the
quasi-classical approximation does not work (the wave length is larger than
the range of a potential inhomogeneity). In particular, this approach has been
applied to the analysis of the time-dependent properties of the scattering by
the imaginary step \cite{Los 2011} (related to the calculation of the particle
time of arrival) and by rectangular barrier/well potentials \cite{Los 2012,
Los 2013}. The GMR and TMR effects in real nanostructures with intermixed
interfaces were considered in \cite{Los and Los (2008)}. Note that our
approach is different from the known MST approach for a layered system in the
stationary case, when, using the reflection and transmission coefficients for
a single potential step (obtained by the traditional matching procedure), the
total reflection and transmission amplitudes for the layered system are then
constructed as the multiple scattering series composed of these known step
coefficients (see, e.g., \cite{Anderson 1989}). This approach, however, says
nothing about a time-dependent picture of a wave scattering by the layered system.

Thus, we show that the mentioned (actually quantum) approach can be directly
extended, e.g., to the light waves propagation in layered systems, i.e. to the
corresponding (second-order in time) wave equation. We show that all
particle/wave scattering situations can be treated in a unified manner by
considering the Green function equation for a corresponding field. The Green
function, related to the wave equation, is obtained through the MST approach
with the interface-localized potentials responsible for a wave reflection from
and transmission through an interface between two media with the different
refraction coefficient. The space-time propagator resolving the wave equation
for the threelayer made of different uniform materials is then expressed
through the found Green function for this threelayer. For the Gaussian wave
packet propagating in this layered system, the solution to the wave equation
in different spatial areas (before, inside and behind the threelayer) is
obtained and numerically visualized as a function of time. It shows, in
particular, the contribution of the backward-moving components of the incoming
Gaussian wave packet to its propagation. \ 

\section{Green function for the wave equation in a layered media}

Let us consider the second-order in time wave equation for a scalar field
(related, e.g., to the light or elastic wave)
\begin{equation}
\left[  -\frac{\partial^{2}}{\partial t^{2}}-H(\mathbf{r})\right]
f(\mathbf{r,}t)=0. \label{1}%
\end{equation}
Here, the function $f(\mathbf{r,}t)$ represents, e.g., any component of the
vector-potential, electric or magnetic field strength and $H(\mathbf{r})$ is a
time-independent, linear, hermitian operator (the Hamiltonian) in the space
representation, which in the abstract vector space can be defined as
\begin{equation}
<\mathbf{r}|H|\mathbf{r}^{\prime}>=\delta(\mathbf{r}-\mathbf{r}^{\prime
})H(\mathbf{r}),f(\mathbf{r,}t)=<\mathbf{r|}f(t)>. \label{2}%
\end{equation}
where $|\mathbf{r>}$ is the eigenvector of the position operator with the
following properties%
\begin{equation}
<\mathbf{r}|\mathbf{r}^{\prime}>=\delta(\mathbf{r}-\mathbf{r}^{\prime}),\int
d\mathbf{r}^{\prime}|\mathbf{r><r}^{\prime}|=1. \label{3}%
\end{equation}

We consider a threelayer made of different uniform materials divided by a
perfect interfaces located at $x=x_{s}$ $s\in\{0,d\}$ planes with $x_{s}=0$
and $x_{d}=d$ (a spacer material of width $d$ is sandwiched between two
semi-infinite layers). We suppose that the operator $H(\mathbf{r})$ can be
split as%
\begin{align}
H(\mathbf{r})  &  =H_{0}(\mathbf{r})+H_{1}(x),\nonumber\\
H_{0}(\mathbf{r})  &  =-v^{2}(x)\mathbf{\nabla}^{2}, \label{4}%
\end{align}
where $H_{1}(x)$ is related to the influence of the interfaces on the wave
free motion in a single layer described by $H_{0}(\mathbf{r})$, and, due to
the system uniformity in the parallel to the interfaces direction, the
perturbation $H_{1}(x)$ and the wave phase velocity $v(x)=\omega/k(x)$
($\omega$ is the wave frequency) both change only in the perpendicular to the
interfaces $x$-direction. Thus,
\begin{align}
v(x)  &  =v_{<}^{0}=v_{1},k(x)=k_{<}^{0}=k_{1}=\omega/v_{1},x<0,\nonumber\\
v(x)  &  =v_{>}^{0}=v_{<}^{d}=v_{2},k(x)=k_{>}^{0}=k_{<}^{d}=k_{2}%
=\omega/v_{2},0<x<d,\nonumber\\
v(x)  &  =v_{>}^{d}=v_{3},k(x)=k_{>}^{d}=k_{3}=\omega/v_{3},x>d, \label{5}%
\end{align}
\qquad where $v_{<(>)}^{s}$ and, correspondingly, $k_{<(>)}^{s}$ define the
phase velocity and the wave number to the left ($<$) and to the right ($>$) of
the interface located at the plane $x=x_{s}$, $s\in\{0,d\}$.

The Green function associated with Eq. (\ref{1}) is defined as the solution of
equation (see, e.g., \cite{Economou})
\begin{equation}
\left[  -\frac{\partial^{2}}{\partial t^{2}}-H(\mathbf{r})\right]
g(\mathbf{r},\mathbf{r}^{\prime};t-t^{\prime})=\delta(\mathbf{r}%
-\mathbf{r}^{\prime})\delta(t-t^{\prime}). \label{6}%
\end{equation}
Expressing the general solution to (\ref{6}) as
\begin{equation}
g(\tau)=\frac{1}{2\pi}%
%TCIMACRO{\dint \limits_{-\infty}^{\infty}}%
%BeginExpansion
{\displaystyle\int\limits_{-\infty}^{\infty}}
%EndExpansion
d\omega e^{-i\omega\tau}G(\omega^{2}), \label{7}%
\end{equation}
we obtain the following equation for $G(\mathbf{r},\mathbf{r}^{\prime}%
;\omega^{2})$%
\begin{equation}
\left[  \omega^{2}-H(\mathbf{r})\right]  G(\mathbf{r},\mathbf{r}^{/}%
;\omega^{2})=\delta(\mathbf{r}-\mathbf{r}^{\prime}). \label{8}%
\end{equation}

For what follows, it is helpful to rewrite Eq. (\ref{8}) in the operator form%
\begin{equation}
G(\mathbf{r},\mathbf{r}^{/};\omega^{2})=<\mathbf{r|}\frac{1}{\omega^{2}%
-H}|\mathbf{r}^{\prime}>. \label{9}%
\end{equation}
We assume on the physical ground that $G(\omega^{2})$ has singularities only
on the real $\omega$-axis. Thus, to make $G(\omega^{2})$ be well defined, the
following limiting procedure is used
\begin{align}
G^{R}(\omega^{2})  &  =\lim_{s\rightarrow+0}G[(\omega+is)^{2}]=\lim
_{s\rightarrow+0}G(\omega^{2}+2i\omega s)=G^{+}(\omega^{2}),\omega
>0,\nonumber\\
G^{R}(\omega^{2})  &  =G^{-}(\omega^{2}),\omega<0,\nonumber\\
G^{A}(\omega^{2})  &  =\lim_{s\rightarrow+0}G[(\omega-is)^{2}]=\lim
_{s\rightarrow+0}G(\omega^{2}-2i\omega s)=G^{-}(\omega^{2}),\omega
>0,\nonumber\\
G^{A}(\omega^{2})  &  =G^{+}(\omega^{2}),\omega<0,\nonumber\\
G^{\pm}(\omega^{2})  &  =\lim_{\varepsilon\rightarrow+0}G(\omega^{2}\pm
i\varepsilon),\varepsilon=2\omega s,\omega>0,s>0. \label{10}%
\end{align}
It is convenient to introduce the function
\begin{align}
\widetilde{g}(\tau)  &  =\frac{1}{2\pi}%
%TCIMACRO{\dint \limits_{-\infty}^{\infty}}%
%BeginExpansion
{\displaystyle\int\limits_{-\infty}^{\infty}}
%EndExpansion
d\omega e^{-i\omega\tau}[G^{R}(\omega^{2})-G^{A}(\omega^{2})]\nonumber\\
&  =-\frac{i}{\pi}%
%TCIMACRO{\dint \limits_{0}^{\infty}}%
%BeginExpansion
{\displaystyle\int\limits_{0}^{\infty}}
%EndExpansion
d\omega\sin(\omega\tau)[G^{+}(\omega^{2})-G^{-}(\omega^{2})], \label{11}%
\end{align}
where Eq. (\ref{10}) is taken into account.

Then, it is not difficult to verify, that the solution to Eq. (\ref{1}) looks
as (see also \cite{Economou})
\begin{equation}
f(\mathbf{r,}t)=-\int d\mathbf{r}^{\prime}\left\{  \widetilde{g}%
(\mathbf{r},\mathbf{r}^{\prime};t-t^{\prime})[\frac{\partial f(\mathbf{r}%
^{\prime}\mathbf{,}t)}{\partial t}]_{t=t^{\prime}}+\frac{\partial\widetilde
{g}(\mathbf{r},\mathbf{r}^{\prime};t-t^{\prime})}{\partial t}f(\mathbf{r}%
^{\prime}\mathbf{,}t^{\prime})\right\}  , \label{12}%
\end{equation}
where
\begin{equation}
\widetilde{g}(\mathbf{r},\mathbf{r}^{\prime};t-t^{\prime})=-\frac{i}{\pi}%
\int\limits_{0}^{\infty}d\omega\sin[\omega(t-t^{\prime})]\left[
G^{+}(\mathbf{r},\mathbf{r}^{/};\omega^{2})-G^{-}(\mathbf{r},\mathbf{r}%
^{/};\omega^{2})\right]  . \label{13}%
\end{equation}

The solution to the Green function $G^{\pm}(\mathbf{r},\mathbf{r}^{\prime
};\omega^{2})$ can be found by means of expanding the Green function (\ref{9})
(the resolvent) into the series in the scattering effective potentials
corresponding to the considered type of the system inhomogeneity (a layered
system). The terms of these series represent the multiple scatterings
processes at the corresponding potentials and their interference. Moreover,
such an approach allows for treating the phenomena under consideration from
the particle-like (electron or photon) rather than the wave point of view. But
in order to realize this program, the effective potentials responsible for
reflection from and transmission through an interface should be found. In
other words, knowing the reflection and transmission coefficients, the inverse
problem of finding the scattering potentials should be solved (see \cite{Los
2010}).

By taking advantage of the two-dimensional uniformity of the system (the
in-plane wave vector $\mathbf{k}_{||}=(k_{y},k_{z})$ is a good quantum
number), the Green function for the perfect interface and the $\delta
$-function may be presented like%
\begin{align}
G^{\pm}(\mathbf{r},\mathbf{r}^{/};\omega^{2})  &  =\frac{1}{A}%
%TCIMACRO{\tsum \limits_{\mathbf{k}_{||}}}%
%BeginExpansion
{\textstyle\sum\limits_{\mathbf{k}_{||}}}
%EndExpansion
e^{i\mathbf{k}_{||}(\mathbf{\rho}-\mathbf{\rho}^{\prime})}G^{\pm}(x,x^{\prime
};\omega^{2};\mathbf{k}_{||}),\nonumber\\
\delta(\mathbf{r}-\mathbf{r}^{\prime})  &  =\frac{\delta(x-x^{\prime})}{A}%
%TCIMACRO{\tsum \limits_{\mathbf{k}_{||}}}%
%BeginExpansion
{\textstyle\sum\limits_{\mathbf{k}_{||}}}
%EndExpansion
e^{i\mathbf{k}_{||}(\mathbf{\rho}-\mathbf{\rho}^{\prime})}, \label{14}%
\end{align}
where $\mathbf{\rho=(}y,z\mathbf{)}$ is a two-dimensional vector in the plane
of the interface and $A$ is the area of the interface. Then, from Eq.
(\ref{8}) we have the one-dimensional equation for the Green function $G^{\pm
}(x,x^{\prime};\omega^{2};\mathbf{k}_{||})$%
\begin{equation}
\left\{  v^{2}(x)[\frac{\partial^{2}}{\partial x^{2}}+k^{\bot2}(x)]-H_{1}%
(x)\right\}  G^{\pm}(x,x^{\prime};\omega^{2};\mathbf{k}_{||})=\delta
(x-x^{\prime}), \label{14a}%
\end{equation}
where%
\begin{equation}
k^{\bot2}(x)=k^{2}(x)-\mathbf{k}_{||}^{2}. \label{15}%
\end{equation}
Thus, the problem is reduced to finding the one-dimensional Green's function
$G^{\pm}(x,x^{\prime};\omega^{2};\mathbf{k}_{||})$ dependent on the conserved
frequency $\omega^{2}$ and the parallel-to-interface component of the wave
vector $\mathbf{k}_{||}$. In the following, we will suppress for simplicity
the dependence on the argument $\mathbf{k}_{||}$, which will be recovered when needed.

The Green functions $G_{0}^{\pm}(x,x^{\prime};\omega^{2})$ corresponding to
the "free" wave propagation in each layer ($x,x^{\prime}<0$, $0<x,x^{\prime
}<d$, $x,x^{\prime}>d$) satisfies, as it follows from Eq. (\ref{14a}), the
following equation%
\begin{equation}
v^{2}(x)[\frac{\partial^{2}}{\partial x^{2}}+k^{\bot2}(x)]G_{0}^{\pm
}(x,x^{\prime};\omega^{2})=\delta(x-x^{\prime}). \label{15a}%
\end{equation}
The Hamiltonian $H_{1}(x)$ describes the perturbations of the "free" motion,
which are, naturally, localized at the steps with coordinates $x_{s}$ (in the
case under consideration, there are two steps at $x_{s}=0$ and $x_{s}=d$).
Thus, we assume that the Hamiltonian $H_{1}(x)$ (generally dependent on
$\omega^{2}$), which corresponds to the processes of scattering at the $v(x)$
steps, can be presented as
\begin{equation}
H_{1}(x;\omega^{2})=%
%TCIMACRO{\dsum \limits_{s}}%
%BeginExpansion
{\displaystyle\sum\limits_{s}}
%EndExpansion
H_{1}^{s}(\omega^{2})\delta(x-x_{s}). \label{16}%
\end{equation}
This perturbation potential depends on the side on which the wave or particle
approaches the interface at $x=x_{s}$ (compare with \cite{Los 2010}, \cite{Los
2012})%

\begin{align}
H_{1>}^{s}(\omega^{2})  &  =iv_{>}^{s2}[k_{>}^{\bot s}(\omega^{2})-k_{<}^{\bot
s}(\omega^{2})],\nonumber\\
H_{1<}^{s}(\omega^{2})  &  =iv_{<}^{s2}[k_{<}^{\bot s}(\omega^{2})-k_{>}^{\bot
s}(\omega^{2})],\nonumber\\
H_{1><}^{s}(\omega^{2})  &  =4iv_{>}^{s}v_{<}^{s}\frac{k_{>}^{\bot s}%
(\omega^{2})k_{<}^{\bot s}(\omega^{2})}{[\sqrt{k_{>}^{\bot s}(\omega^{2}%
)}+\sqrt{k_{<}^{\bot s}(\omega^{2})}]^{2}}. \label{17}%
\end{align}
Here, $H_{1i>(<)}^{s}(\omega^{2})$ is the reflection (from the potential step
at $x=x_{s}$, $s\in\{0,d\}$) potential amplitude, the index $>(<)$ indicates
the side on which the wave approaches the interface at $x=x_{s}$: right ($>$)
or left ($<$); $H_{i><}^{s}(\omega^{2})$ is the transmission potential
amplitude. More exactly, the perturbation $H_{1}^{s}$ depends on the parameter
$\omega^{2}\pm i\varepsilon$ (for the $G^{\pm}$ Green function) and
$\mathbf{k}_{||}$ (which is omitted for brevity) via the
perpendicular-to-interfaces wave vector (\ref{6})%
\begin{equation}
k_{>(<)}^{\bot s}(\omega^{2})=\sqrt{(k_{>(<)}^{s})^{2}-\mathbf{k}_{||}^{2}}
\label{18}%
\end{equation}
where $k_{>(<)}^{s}$ ($s\in\{0,d\}$) are defined by Eqs. (\ref{5}), and,
correspondingly, $v_{>(<)}^{s}$ takes the values $v_{1}$, $v_{2}$ or $v_{3}$.

The perturbation expansion for the retarded Green function $G^{+}(x,x^{\prime
};\omega^{2})$ in the case of the two-step effective scattering potential
(\ref{16}), reads for different areas of arguments $x$, $x^{\prime}$ as (see
\cite{Los 2012,Los 2013})%
\begin{align}
G^{+}(x,x^{\prime};\omega^{2})  &  =G_{0}^{+}(x,d;\omega^{2})T^{+}(\omega
^{2})G_{0}^{+}(0,x^{\prime};\omega^{2}),x^{\prime}<0,x>d,\nonumber\\
G^{+}(x,x^{\prime};\omega^{2})  &  =G_{0}^{+}(x,0;\omega^{2})T^{+}(\omega
^{2})G_{0}^{+}(d,x^{\prime};\omega^{2}),x^{\prime}>d,x<0,\nonumber\\
G^{+}(x,x^{\prime};\omega^{2})  &  =G_{0}^{+}(x,0;\omega^{2})T^{\prime
+}(\omega^{2})G_{0}^{+}(0,x^{\prime};\omega^{2})+G_{0}^{+}(x,d;\omega
^{2})R^{\prime+}(\omega^{2})G_{0}^{+}(0,x^{\prime};\omega^{2}),x^{/}%
<0,0<x<d,\nonumber\\
G^{+}(x,x^{\prime};\omega^{2})  &  =G_{0}^{+}(x,0;\omega^{2})T^{\prime
+}(\omega^{2})G_{0}^{+}(0,x^{\prime};\omega^{2})+G_{0}^{+}(x,0;\omega
^{2})R^{\prime+}(\omega^{2})G_{0}^{+}(d,x^{\prime};\omega^{2}),0<x^{/}%
<d,x<0,\nonumber\\
G^{+}(x,x^{\prime};\omega^{2})  &  =G_{0}^{+}(x,x^{\prime};\omega^{2}%
)+G_{0}^{+}(x,0;\omega^{2})R^{+}(\omega^{2})G_{0}^{+}(0,x^{\prime};\omega
^{2}),x^{\prime}<0,x<0, \label{18a}%
\end{align}
and the transmission and reflection amplitudes are
\begin{align}
T^{+}(\omega^{2})  &  =\frac{T_{><}^{d+}(\omega^{2})G_{0}^{+}(d,0;\omega
^{2})T_{><}^{0+}(\omega^{2})}{D^{+}(\omega^{2})},\nonumber\\
T^{\prime+}(\omega^{2})  &  =\frac{T_{><}^{0+}(\omega^{2})}{D^{+}(\omega^{2}%
)},R^{\prime+}(\omega^{2})=T_{<}^{d+}(\omega^{2})G_{0}^{+}(d,0;\omega
^{2})T^{\prime+}(\omega^{2}),\nonumber\\
R^{+}(\omega^{2})  &  =T_{<}^{0+}(\omega^{2})+\frac{T_{><}^{0+}(\omega
^{2})G_{0}^{+}(0,d;\omega^{2})T_{<}^{d+}(\omega^{2})G_{0}^{+}(d,0;\omega
^{2})T_{><}^{0+}(\omega^{2})}{D^{+}(\omega^{2})}\nonumber\\
D^{+}(\omega^{2})  &  =1-G_{0}^{+}(d,0;\omega^{2})T_{>}^{0+}(\omega^{2}%
)G_{0}^{+}(0,d;\omega^{2})T_{<}^{d+}(\omega^{2}). \label{19}%
\end{align}
The one-dimensional retarded Green function corresponding to a "free" motion
in an uniform media of some layer, $G_{0}^{+}(x,x^{\prime};\omega^{2})$, is
(see, e.g. \cite{Economou})%
\begin{equation}
G_{0}^{+}(x,x^{\prime};\omega^{2})=\frac{1}{i2v_{>(<)}^{s2}k_{>(<)}^{\bot
s}(\omega^{2})}\exp[ik_{>(<)}^{\bot s}(\omega^{2})|x-x^{\prime}|],s=0,d.
\label{20}%
\end{equation}
The reflection $T_{>(<)}^{s+}(\omega^{2})$ and transmission $T_{><}%
^{s+}(\omega^{2})$ amplitude, used in (\ref{19}), corresponding to the
retarded Green function and scattering at the interface located at $x=x_{s}%
\in\{0,d\}$, are defined by the following perturbation expansion:%
\begin{align}
T^{s}(\omega^{2})  &  =H_{1}^{s}(\omega^{2})+H_{1}^{s}(\omega^{2})G_{0}%
(x_{s},x_{s};\omega^{2})H_{1}^{s}(\omega^{2})+\ldots\nonumber\\
&  =\frac{H_{1}^{s}(\omega^{2})}{1-G_{0}(x_{s},x_{s};\omega^{2})H_{1}%
^{s}(\omega^{2})}, \label{21}%
\end{align}
where $H_{i}^{s}(\omega^{2})$ and the interface Green function $G_{0}%
(x_{s},x_{s};\omega^{2})$ are defined differently for reflection and
transmission processes \cite{Los 2010, Los 2012}: the step-localized effective
potential is given by Eq. (\ref{17}) and the retarded Green functions at the
interface for the considered reflection and transmission processes are,
correspondingly,
\begin{align}
G_{0>(<)}^{+}(x_{s},x_{s};\omega^{2})  &  =1/2iv_{>(<)}^{s2}k_{>(<)}^{\bot
s}(\omega^{2}),\nonumber\\
G_{0><}^{+}(x_{s},x_{s};\omega^{2})  &  =1/2iv_{>}^{s}v_{<}^{s}\sqrt
{k_{>}^{\bot s}(\omega^{2})k_{<}^{\bot s}(\omega^{2})} \label{22}%
\end{align}
in accordance with (\ref{20}).

From \ref{17}), (\ref{21}) and (\ref{22}), we have for $T_{>(<)}^{s+}(E)$ and
$T_{><}^{s+}(E)$
\begin{align}
T_{>(<)}^{s+}(\omega^{2})  &  =2iv_{>(<)}^{s2}k_{>(<)}^{\bot s}r_{>(<)}%
^{s},\nonumber\\
T_{><}^{s+}(\omega^{2})  &  =2iv_{>}^{s}v_{<}^{s}\sqrt{k_{>}^{\bot s}%
k_{<}^{\bot s}}t^{s}, \label{23}%
\end{align}
where $r_{>(<)}^{s}(\lambda)$ and $t^{s}(\lambda)$ are the standard amplitudes
for reflection to the right (left) of the potential step at $x=x_{s}$ and
transmission through this step%
\begin{align}
r_{>}^{s}(\omega^{2})  &  =\frac{k_{>}^{\bot s}-k_{<}^{\bot s}}{k_{>}^{\bot
s}+k_{<}^{\bot s}},r_{<}^{s}(\omega^{2})=\frac{k_{<}^{\bot s}-k_{>}^{\bot s}%
}{k_{>}^{\bot s}+k_{<}^{\bot s}},\nonumber\\
t^{s}(\omega^{2})  &  =\frac{2\sqrt{k_{>}^{\bot s}k_{<}^{\bot s}}}{k_{>}^{\bot
s}+k_{<}^{\bot s}}, \label{24}%
\end{align}
and the argument $\omega^{2}$ in the wave vectors is omitted for brevity.

The expressions (\ref{24}) can be rewritten in the form of the Fresnel
formulae (e.g. for the wave incoming from the right to an interface with index
$s$)
\begin{align}
r_{>}^{s}(\omega^{2})  &  =\frac{n_{><}^{s}\cos\varphi_{>}^{s}-\cos\varphi
_{<}^{s}}{n_{><}^{s}\cos\varphi_{>}^{s}+\cos\varphi_{<}^{s}},\nonumber\\
t^{s}(\omega^{2})  &  =\frac{2\sqrt{n_{><}^{s}\cos\varphi_{>}^{s}\cos
\varphi_{<}^{s}}}{n_{><}^{s}\cos\varphi_{>}^{s}+\cos\varphi_{<}^{s}}.
\label{25}%
\end{align}
Here $\cos\varphi_{>(<)}^{s}=k_{>(<)}^{\bot s}/k_{>(<)}^{s}$, i.e.
$\varphi_{>(<)}^{s}$ is the angle of incidence at the interface from right
(left), $n_{><}^{s}$ is the relative refraction coefficient
\begin{equation}
n_{><}^{s}=\frac{k_{>}^{s}}{k_{<}^{s}}. \label{26}%
\end{equation}
Note, that for the light waves the expressions (\ref{25}) correspond to the
case when the electrical strength vector is perpendicular to the plane of
incidence (in this case the wave polarization does not change at reflection
and refraction).

Using Eqs. (\ref{5}), (\ref{15}), (\ref{18a}), (\ref{19}), (\ref{20}),
(\ref{23}) and (\ref{24}), we obtain%
\begin{align}
G^{+}(x,x^{\prime};\omega^{2})  &  =\frac{1}{i2v_{1}v_{3}\sqrt{k_{1}^{\bot
}k_{3}^{\bot}}}e^{ik_{3}^{\bot}(x-d)}t(\omega^{2})e^{-ik_{1}^{\bot}x^{\prime}%
},x^{\prime}<0,x>d,\nonumber\\
G^{+}(x,x^{\prime};\omega^{2})  &  =\frac{1}{i2v_{1}v_{3}\sqrt{k_{1}^{\bot
}k_{3}^{\bot}}}e^{-ik_{1}^{\bot}x}t(\omega^{2})e^{ik_{3}^{\bot}(x^{\prime}%
-d)},x^{\prime}>d,x<0,\nonumber\\
G^{+}(x,x^{\prime};\omega^{2})  &  =\frac{1}{i2v_{1}v_{2}\sqrt{k_{1}^{\bot
}k_{2}^{\bot}}}\left[  e^{ik_{2}^{\bot}x}t^{\prime}(\omega^{2})e^{-ik_{1}%
^{\bot}x^{\prime}}+e^{-ik_{2}^{\bot}x}r^{\prime}(\omega^{2})e^{-ik_{1}^{\bot
}x^{\prime}}\right]  ,x^{\prime}<0,0<x<d,\nonumber\\
G^{+}(x,x^{\prime};\omega^{2})  &  =\frac{1}{i2v_{1}v_{2}\sqrt{k_{1}^{\bot
}k_{2}^{\bot}}}\left[  e^{-ik_{1}^{\bot}x}t^{\prime}(\omega^{2})e^{ik_{2}%
^{\bot}x^{\prime}}+e^{-ik_{1}^{\bot}x}r^{\prime}(\omega^{2})e^{-ik_{2}^{\bot
}x^{\prime}}\right]  ,x<0,0<x^{\prime}<d,\nonumber\\
G^{+}(x,x^{\prime};\omega^{2})  &  =\frac{1}{i2v_{1}^{2}k_{1}^{\bot}}\left[
e^{ik_{1}^{\bot}\left\vert x-x^{\prime}\right\vert }+r(\omega^{2}%
)e^{-ik_{1}^{\bot}(x+x^{\prime})}\right]  ,x<0,x^{\prime}<0, \label{27}%
\end{align}
where the transmission and reflection amplitudes are defined as%
\begin{align}
t(\omega^{2})  &  =\frac{4\sqrt{k_{1}^{\bot}k_{3}^{\bot}}k_{2}^{\bot}%
e^{ik_{2}^{\bot}d}}{d(\omega^{2})},t^{\prime}(\omega^{2})=\frac{2\sqrt
{k_{1}^{\bot}k_{2}^{\bot}}(k_{3}^{\bot}+k_{2}^{\bot})}{d(\omega^{2}%
)},\nonumber\\
r^{\prime}(\omega^{2})  &  =\frac{2\sqrt{k_{1}^{\bot}k_{2}^{\bot}}(k_{2}%
^{\bot}-k_{3}^{\bot})e^{2ik_{2}^{\bot}d}}{d(\omega^{2})},r(\omega^{2}%
)=\frac{(k_{1}^{\bot}-k_{2}^{\bot})(k_{3}^{\bot}+k_{2}^{\bot})-(k_{1}^{\bot
}+k_{2}^{\bot})(k_{3}^{\bot}-k_{2}^{\bot})e^{2ik_{2}^{\bot}d}}{d(\omega^{2}%
)},\nonumber\\
d(\omega^{2})  &  =(k_{1}^{\bot}+k_{2}^{\bot})(k_{3}^{\bot}+k_{2}^{\bot
})-(k_{1}^{\bot}-k_{2}^{\bot})(k_{3}^{\bot}-k_{2}^{\bot})e^{2ik_{2}^{\bot}d},
\label{28}%
\end{align}
where the perpendicular-to-interface wave vectors $k_{i}^{\bot}$ ($i=1,2,3$)
are defined by (\ref{15}) and (\ref{5})%
\begin{equation}
k_{i}^{\bot}=\sqrt{k_{i}^{2}-\mathbf{k}_{||}^{2}}=\sqrt{\frac{\omega^{2}%
}{v_{i}^{2}}-k_{||}^{2}},i=1,2,3. \label{28a}%
\end{equation}
From the last line of Eq. (\ref{27}) one can see that the "free" Green
function in any layer is (see also \cite{Economou})
\begin{equation}
G_{0}^{+}(x,x^{\prime};\omega^{2};\mathbf{k}_{||}^{2})=\frac{1}{i2v_{i}%
^{2}k_{i}^{\bot}}e^{ik_{i}^{\bot}\left\vert x-x^{\prime}\right\vert }
\label{28b}%
\end{equation}
in accordance with (\ref{20}). Obtained result, given by Eqs. (\ref{27}) and
(\ref{28}), allows for finding the solution to the second-order in time (wave)
equation. The corresponding space-time propagator is defined by Eqs.
(\ref{13}), (\ref{14}), and the solution of the wave equation can be found
from (\ref{12}).

The transmission probability $\left\vert t(\omega^{2})\right\vert ^{2}$
through and reflection probability $\left\vert r(\omega^{2})\right\vert ^{2} $
from the asymmetric threelayer ($v_{1}\neq v_{3}$), which follow from
(\ref{28}), for real $k_{2}^{\bot}$ and $k_{3}^{\bot}$, are given by
\begin{align}
\left\vert t(\omega^{2})\right\vert ^{2}  &  =\frac{4k_{1}^{\bot}k_{2}^{\bot
2}k_{3}^{\bot}}{(k_{1}^{\bot}+k_{3}^{\bot})^{2}k_{2}^{\bot2}+(k_{1}^{\bot
2}-k_{2}^{\bot2})(k_{3}^{\bot2}-k_{2}^{\bot2})\sin^{2}(k_{2}^{\bot}%
d)},\nonumber\\
\left\vert r(\omega^{2})\right\vert ^{2}  &  =\frac{k_{2}^{\bot2}(k_{1}^{\bot
}-k_{3}^{\bot})^{2}+(k_{1}^{\bot2}-k_{2}^{\bot2})(k_{3}^{\bot2}-k_{2}^{\bot
2})\sin^{2}(k_{2}^{\bot}d)}{(k_{1}^{\bot}+k_{3}^{\bot})^{2}k_{2}^{\bot
2}+(k_{1}^{\bot2}-k_{2}^{\bot2})(k_{3}^{\bot2}-k_{2}^{\bot2})\sin^{2}%
(k_{2}^{\bot}d)}. \label{29}%
\end{align}
Note that when $k_{2}^{\bot}d=n\pi$ ($n$ is integer), the resonance
transmission ($\left\vert t(\omega^{2})\right\vert ^{2}=1$ and $\left\vert
r(\omega^{2})\right\vert ^{2}=0$) happens only for a symmetric layered system
($k_{3}^{\bot}=k_{1}^{\bot}$).

From Eqs. (\ref{27}) we see that $G^{+}(x,x^{\prime};\omega^{2})=G^{+}%
(x^{\prime},x;\omega^{2})$, and, therefore, the advanced Green function
$G^{-}(x,x^{\prime};\omega^{2})=\left[  G^{+}(x^{\prime},x;\omega^{2})\right]
^{\ast}=\left[  G^{+}(x,x^{\prime};\omega^{2})\right]  ^{\ast} $ (see, e.g.
\cite{Economou}). Thus, the transmission amplitude (\ref{13}) (propagator) is
determined by the imaginary part of the Green function (\ref{27}) and with
account of (\ref{14}) can be written as%
\begin{align}
\widetilde{g}(\mathbf{r},\mathbf{r}^{\prime};t-t^{\prime})  &  =\frac{1}{A}%
%TCIMACRO{\tsum \limits_{\mathbf{k}_{||}}}%
%BeginExpansion
{\textstyle\sum\limits_{\mathbf{k}_{||}}}
%EndExpansion
e^{i\mathbf{k}_{||}(\mathbf{\rho}-\mathbf{\rho}^{\prime})}\widetilde
{g}(x,x^{\prime};t-t^{\prime};\mathbf{k}_{||}),\nonumber\\
\widetilde{g}(x,x^{\prime};t-t^{\prime};\mathbf{k}_{||})  &  =-\frac{i}{\pi
}\int\limits_{0}^{\infty}d\omega\sin[\omega(t-t^{\prime})]\left[
G^{+}(x,x^{/};\omega^{2};\mathbf{k}_{||})-G^{-}(x,x^{/};\omega^{2}%
;\mathbf{k}_{||})\right] \nonumber\\
&  =\frac{2}{\pi}\int\limits_{0}^{\infty}d\omega\sin[\omega(t-t^{\prime
})]\operatorname{Im}G^{+}(x,x^{/};\omega^{2};\mathbf{k}_{||}). \label{30}%
\end{align}
Here, the Green functions (\ref{27}) depend on $\mathbf{k}_{||}$ via
$k_{i}^{\bot}$ (\ref{28a}) and these functions correspond to the fixed angle
of the wave incidence at the interface defined by $\mathbf{k}_{||}$. Note,
that for the incoming-to-interface real wave from the layer $1$ ($x<0$), the
integration over $\omega$ in (\ref{30}) should start with $\omega=v_{1}k_{||}$
(see (\ref{28a}), (\ref{28b})).

Eqations (\ref{12}), (\ref{27}) and (\ref{30}) give the complete exact
solution of the wave equation (\ref{1}) for \ the layered system at any given
initial value $f(\mathbf{r}^{\prime},t^{\prime})$.

\section{Solution to wave equation and numerical modeling}

\bigskip In order to consider the wave propagation in time, we will adopt for
Eq. (\ref{12}) the initial (at $t^{\prime}=0$) wave in the form of the
Gaussian wave packet localized at the $x<0$ area
\begin{equation}
f(\mathbf{r}^{\prime},t^{\prime}=0)=C\exp\left[  -\frac{(x^{\prime}-x_{i}%
)^{2}}{2\sigma_{x}^{2}}\right]  \exp\left[  -\frac{(\mathbf{\rho}^{\prime
}-\mathbf{\rho}_{i})^{2}}{2\sigma_{||}^{2}}\right]  \cos(\mathbf{k}%
^{0}\mathbf{r}^{\prime}),x_{i}<0,k_{x}^{0}>0, \label{31}%
\end{equation}
where $C$ is a constant (e.g., the strength of the electric field $E_{y}$),
$\sigma_{x}^{2}$ and $\sigma_{||}^{2}$ are the dispersions in the
perpendicular-to- and parallei-to-interface directions, $\mathbf{r}^{\prime
}=(x^{\prime},\mathbf{\rho}^{\prime})$, $\mathbf{k}^{0}=(k_{x}^{0}%
,\mathbf{k}_{||}^{0})$, $\left\vert \mathbf{k}^{0}\right\vert =\omega
^{0}/v_{1}$, $\omega^{0}$ is the frequency of the modulated incoming wave.
Thus, we consider a situation, when the wave packet (\ref{31}), located in the
vicinity of $\mathbf{r}_{i}=(x_{i},\mathbf{\rho}_{i})$, comes from the left
($x<0$) with the positive perpendicular-to-interface component of the wave
vector $k_{x}^{0}>0$ at the angle defined by the parallel-to-inteface wave
vector component $\mathbf{k}_{||}^{0}$. Substituting (\ref{31}) into
(\ref{12}) and using (\ref{30}), we can present the solution as
\begin{align}
f(\mathbf{r},t)  &  =f_{+}(\mathbf{r},t)+f_{-}(\mathbf{r},t),\nonumber\\
f_{\pm}(\mathbf{r},t)  &  =-\frac{C}{\pi}%
%TCIMACRO{\dint \limits_{v_{1}k_{||}}^{\infty}}%
%BeginExpansion
{\displaystyle\int\limits_{v_{1}k_{||}}^{\infty}}
%EndExpansion
d\omega\omega\int d\mathbf{r}^{\prime}\frac{1}{A}%
%TCIMACRO{\dsum \limits_{\mathbf{k}_{||}}}%
%BeginExpansion
{\displaystyle\sum\limits_{\mathbf{k}_{||}}}
%EndExpansion
e^{i\mathbf{k}_{||}(\mathbf{\rho}-\mathbf{\rho}^{\prime})}\exp\left[
-\frac{(x^{\prime}-x_{i})^{2}}{2\sigma_{x}^{2}}\right]  \exp\left[
-\frac{(\mathbf{\rho}^{\prime}-\mathbf{\rho}_{i})^{2}}{2\sigma_{||}^{2}%
}\right] \nonumber\\
&  \times\operatorname{Im}[G^{+}(x,x^{\prime};\omega^{2};\mathbf{k}_{||}%
)]\cos(\omega t\mp\mathbf{k}^{0}\mathbf{r}^{\prime}), \label{31a}%
\end{align}
where $f_{+}(\mathbf{r},t)$ can be interpreted as a wave moving in the
positive direction of $x$-axis, while $f_{-}(\mathbf{r},t)$ corresponds to the
moving in the opposite direction. Integrating over spatial variables
$\mathbf{\rho}^{\prime}$ ($y^{\prime},z^{\prime}$), we get
\begin{align}
f_{\pm}(\mathbf{r},t)  &  =-2C\int\limits_{v_{1}k_{||}}^{\infty}d\omega\omega%
%TCIMACRO{\dint \limits_{-\infty}^{\infty}}%
%BeginExpansion
{\displaystyle\int\limits_{-\infty}^{\infty}}
%EndExpansion
dx^{\prime}e^{-(x^{\prime}-x_{i})^{2}/2\sigma_{x}^{2}}\nonumber\\
&  \times\frac{1}{A}%
%TCIMACRO{\dsum \limits_{\mathbf{k}_{||}}}%
%BeginExpansion
{\displaystyle\sum\limits_{\mathbf{k}_{||}}}
%EndExpansion
\sigma_{||}^{2}e^{-\frac{(\mathbf{k}_{||}^{0}-\mathbf{k}_{||})^{2}}{2}%
\sigma_{||}^{2}}\left[  \operatorname{Im}G^{+}(x,x^{\prime};\omega
^{2};\mathbf{k}_{||})\right]  \cos\{\omega t\mp\lbrack k_{x}^{0}x^{\prime
}+(\mathbf{k}_{||}^{0}-\mathbf{k}_{||})\mathbf{\rho}_{i}+\mathbf{k}%
_{||}\mathbf{\rho}]\}, \label{32}%
\end{align}
where the property $G^{+}(x,x^{\prime};\omega^{2};\mathbf{k}_{||}%
)=G^{+}(x,x^{\prime};\omega^{2};-\mathbf{k}_{||})$ is also used (see
(\ref{14a}), (\ref{15})). It is not difficult to integrate (\ref{32}) over
$x^{\prime}$. Using for $G^{+}(x,x^{\prime};\omega^{2};\mathbf{k}_{||})$ Eqs.
(\ref{27}), this integration yields
\begin{align}
f_{+}(\mathbf{r,}t)  &  =-C\sqrt{2\pi}\sigma_{x}\int\limits_{v_{1}k_{||}%
}^{\infty}d\omega\omega\frac{1}{A}%
%TCIMACRO{\dsum \limits_{\mathbf{k}_{||}}}%
%BeginExpansion
{\displaystyle\sum\limits_{\mathbf{k}_{||}}}
%EndExpansion
\sigma_{||}^{2}e^{-\frac{(\mathbf{k}_{||}^{0}-\mathbf{k}_{||})^{2}}{2}%
\sigma_{||}^{2}}\nonumber\\
&  \times\operatorname{Im}\{G^{+}(x,x_{i};\omega^{2};\mathbf{k}_{||}%
)\exp\left(  i[\omega t-k_{x}^{0}x_{i}-(\mathbf{k}_{||}^{0}-\mathbf{k}%
_{||})\mathbf{\rho}_{i}-\mathbf{k}_{||}\mathbf{\rho}]\right)  e^{-(k_{1}%
^{\bot}+k_{x}^{0})^{2}\sigma_{x}^{2}/2}\nonumber\\
&  +G^{+}(x,x_{i};\omega^{2};\mathbf{k}_{||})\exp\left(  -i[\omega t-k_{x}%
^{0}x_{i}-(\mathbf{k}_{||}^{0}-\mathbf{k}_{||})\mathbf{\rho}_{i}%
-\mathbf{k}_{||}\mathbf{\rho}]\right)  e^{-(k_{1}^{\bot}-k_{x}^{0})^{2}%
\sigma_{x}^{2}/2}\},\nonumber\\
f_{-}(\mathbf{r,}t)  &  =-C\sqrt{2\pi}\sigma_{x}\int\limits_{v_{1}k_{||}%
}^{\infty}d\omega\omega\frac{1}{A}%
%TCIMACRO{\dsum \limits_{\mathbf{k}_{||}}}%
%BeginExpansion
{\displaystyle\sum\limits_{\mathbf{k}_{||}}}
%EndExpansion
\sigma_{||}^{2}e^{-\frac{(\mathbf{k}_{||}^{0}-\mathbf{k}_{||})^{2}}{2}%
\sigma_{||}^{2}}\nonumber\\
&  \times\operatorname{Im}\{G^{+}(x,x_{i};\omega^{2};\mathbf{k}_{||}%
)\exp\left(  i[\omega t+k_{x}^{0}x_{i}+(\mathbf{k}_{||}^{0}-\mathbf{k}%
_{||})\mathbf{\rho}_{i}+\mathbf{k}_{||}\mathbf{\rho}]\right)  e^{-(k_{1}%
^{\bot}-k_{x}^{0})^{2}\sigma_{x}^{2}/2}\nonumber\\
&  +G^{+}(x,x_{i};\omega^{2};\mathbf{k}_{||})\exp\left(  -i[\omega t+k_{x}%
^{0}x_{i}+(\mathbf{k}_{||}^{0}-\mathbf{k}_{||})\mathbf{\rho}_{i}%
+\mathbf{k}_{||}\mathbf{\rho}]\right)  e^{-(k_{1}^{\bot}+k_{x}^{0})^{2}%
\sigma_{x}^{2}/2}\} \label{32a}%
\end{align}
Here, the Green function $G^{+}(x,x_{i};\omega^{2};\mathbf{k}_{||})$ is
defined in the different spatial areas by Eq. (\ref{27}) with $x^{\prime}$
replaced with $x_{i}$. Expressions (\ref{32a}) provide the exact solution for
the wave equation in a layered system in the case of the initial Gaussian wave
packet (\ref{31}).

We focus on the processes of the transmission through and reflection from a
spacer ($0<x<d$) in time. With this in mind, we further simplify Eqs.
(\ref{32a}) by considering the case of very large dispersion in the
parallel-to-interface plane ($\sigma_{||}\rightarrow\infty$) using the
formula
\begin{equation}
\lim_{\sigma_{||}\rightarrow\infty}\frac{1}{2\pi}\sigma_{||}^{2}%
e^{-(\mathbf{k}_{||}^{0}-\mathbf{k}_{||})^{2}\sigma_{||}^{2}/2}=\delta
(\mathbf{k}_{||}^{0}-\mathbf{k}_{||}). \label{33}%
\end{equation}
Now, integrating (\ref{32a}) over $\mathbf{k}_{||}$, we obtain%
\begin{align}
f_{\pm}(\mathbf{r,}t)  &  =-\frac{C}{\sqrt{2\pi}}\int\limits_{v_{1}k_{||}^{0}%
}^{\infty}d\omega\omega\operatorname{Im}[e^{\mp i\omega t}\Phi(\mathbf{r,}%
x_{i};\omega;\mathbf{k}^{0})+e^{\pm i\omega t}\Phi(\mathbf{r,}x_{i}%
;\omega;-\mathbf{k}^{0})]\nonumber\\
\Phi(\mathbf{r,}x_{i};\omega;\mathbf{k}^{0})  &  =\sigma_{x}e^{-(k_{1}^{\bot
0}-k_{x}^{0})^{2}\sigma_{x}^{2}/2}G^{+}(x,x_{i};\omega^{2};\mathbf{k}_{||}%
^{0})e^{i(k_{x}^{0}x_{i}+\mathbf{k}_{||}^{0}\mathbf{\rho})},k_{1}^{\bot
0}=\sqrt{k_{1}^{2}-\mathbf{k}_{||}^{0^{2}}}. \label{34}%
\end{align}
At last, for a normal (perpendicular-to-interface) incoming wave packet
($\mathbf{k}_{||}^{0}=0$)%
\begin{align}
f_{\pm}(x,t)  &  =-\frac{C}{\sqrt{2\pi}}\int\limits_{0}^{\infty}d\omega
\omega\operatorname{Im}[e^{\mp i\omega t}\Phi(x,x_{i};\omega;k_{x}^{0})+e^{\pm
i\omega t}\Phi(x,x_{i};\omega;-k_{x}^{0})],\nonumber\\
\Phi(x,x_{i};\omega;k_{x}^{0})  &  =\sigma_{x}\exp[-(k_{x}^{0}-k_{1}%
)^{2}\sigma_{x}^{2}/2]G^{+}(x,x_{i};\omega)e^{ik_{x}^{0}x_{i}}. \label{35}%
\end{align}
where, e.g., for the transmission through a spacer case%
\begin{align}
G^{+}(x,x_{i};\omega)  &  =\frac{1}{i2v_{1}v_{3}\sqrt{k_{1}k_{3}}}%
e^{ik_{3}(x-d)}t(\omega)e^{-ik_{1}x_{i}},x_{i}<0,x>d\nonumber\\
t(\omega)  &  =\frac{4\sqrt{k_{1}k_{3}}k_{2}e^{ik_{2}d}}{d(\omega)}%
,k_{i}=\frac{\omega}{v_{i}},i=1,2,3,\nonumber\\
d(\omega)  &  =(k_{1}+k_{2})(k_{3}+k_{2})-(k_{1}-k_{2})(k_{3}-k_{2}%
)e^{2ik_{2}d}. \label{36}%
\end{align}

For what follows and numerical evaluation, let us rewrite Eq. (\ref{35}) in
the dimensionless variables. There is a natural distance unit $d$ (the spacer
width) and for the frequency (time) unit we choose $\omega_{d}=v_{2}/d$
($t_{d}=d/v_{2}$), where $v_{2}$ is the phase velocity in the $0<x<d$ area
(see (\ref{5})). The result is%
\begin{equation}
f_{\pm}(\widetilde{x},\widetilde{t})=-\frac{C}{\sqrt{2\pi}}\int\limits_{0}%
^{\infty}d\widetilde{\omega}\operatorname{Im}[e^{\mp i\widetilde{\omega
}\widetilde{t}}\varphi(\widetilde{x},\widetilde{x}_{i};\widetilde{\omega
};\widetilde{\omega}^{0})+e^{\pm i\widetilde{\omega}\widetilde{t}}%
\varphi(\widetilde{x},\widetilde{x}_{i};\widetilde{\omega};-\widetilde{\omega
}^{0})], \label{40}%
\end{equation}
where%
\begin{align}
\varphi(\widetilde{x},\widetilde{x}_{i};\widetilde{\omega};\widetilde{\omega
}^{0})  &  =-2i\frac{v_{2}}{v_{1}}\frac{v_{2}}{v_{3}}\widetilde{\sigma}%
_{x}\exp[-\frac{v_{2}^{2}}{v_{1}^{2}}(\widetilde{\omega}^{0}-\widetilde
{\omega})^{2}\widetilde{\sigma}_{x}^{2}/2]e^{i\widetilde{\omega}}%
e^{i\frac{v_{2}}{v_{3}}\widetilde{\omega}(\widetilde{x}-1)}\nonumber\\
\times e^{i\frac{v_{2}}{v_{1}}(\widetilde{\omega}^{0}-\widetilde{\omega
})\widetilde{x}_{i}}\frac{1}{\widetilde{d}(\widetilde{\omega})},\widetilde
{x}_{i}  &  <0,\widetilde{x}>1,\nonumber\\
\widetilde{d}(\widetilde{\omega})  &  =(\frac{v_{2}}{v_{1}}+1)(\frac{v_{2}%
}{v_{3}}+1)-(\frac{v_{2}}{v_{1}}-1)(\frac{v_{2}}{v_{3}}-1)e^{2i\widetilde
{\omega}},\nonumber\\
\varphi(\widetilde{x},\widetilde{x}_{i};\widetilde{\omega};\widetilde{\omega
}^{0})  &  =-i\frac{v_{2}}{v_{1}}\widetilde{\sigma}_{x}\exp[-\frac{v_{2}^{2}%
}{v_{1}^{2}}(\widetilde{\omega}^{0}-\widetilde{\omega})^{2}\widetilde{\sigma
}_{x}^{2}/2][e^{i\widetilde{\omega}\widetilde{x}}(\frac{v_{2}}{v_{3}%
}+1)\nonumber\\
+e^{-i\widetilde{\omega}\widetilde{x}}(1-\frac{v_{2}}{v_{3}})e^{2i\widetilde
{\omega}}]\exp[i\frac{v_{2}}{v_{1}}(\widetilde{\omega}^{0}-\widetilde{\omega
})\widetilde{x}_{i}]\frac{1}{\widetilde{d}(\widetilde{\omega})},\widetilde
{x}_{i}  &  <0,0<\widetilde{x}<1,\nonumber\\
\varphi(\widetilde{x},\widetilde{x}_{i};\widetilde{\omega};\widetilde{\omega
}^{0})  &  =-\frac{i}{2}\frac{v_{2}}{v_{1}}\widetilde{\sigma}_{x}\exp
[-\frac{v_{2}^{2}}{v_{1}^{2}}(\widetilde{\omega}^{0}-\widetilde{\omega}%
)^{2}\widetilde{\sigma}_{x}^{2}/2][e^{i\frac{v_{2}}{v_{1}}\widetilde{\omega
}\widetilde{x}}\nonumber\\
+\widetilde{r}(\widetilde{\omega})e^{-i\frac{v_{2}}{v_{1}}\widetilde{\omega
}\widetilde{x}}]e^{i\frac{v_{2}}{v_{1}}(\widetilde{\omega}^{0}-\widetilde
{\omega})\widetilde{x}_{i}},\widetilde{x}_{i}  &  <0,\widetilde{x}%
<0,\nonumber\\
\widetilde{r}(\widetilde{\omega})  &  =[(\frac{v_{2}}{v_{1}}-1)(\frac{v_{2}%
}{v_{3}}+1)-(\frac{v_{2}}{v_{1}}+1)(\frac{v_{2}}{v_{3}}-1)e^{2i\widetilde
{\omega}}]/\widetilde{d}(\widetilde{\omega}) \label{41}%
\end{align}
Here, the dimensionless variables and parameters are%
\begin{equation}
\widetilde{x}=x/d,\widetilde{x}_{i}=x_{i}/d,\widetilde{\sigma}_{x}=\sigma
_{x}/d,\widetilde{\omega}=\omega/\omega_{d},\widetilde{\omega}^{0}=\omega
^{0}/\omega_{d},\widetilde{t}=t/t_{d},\omega_{d}=v_{2}/d,t_{d}=\omega_{d}%
^{-1}. \label{42}%
\end{equation}

As one can see from Eq. (\ref{40}), for an incoming wave packet (\ref{31}) of
the finite width $\widetilde{\sigma}_{x}$(in the $x$ direction), the plane
waves making up this wave packet and moving in both directions along $x$-axis
contribute to the solution $f_{\pm}(\widetilde{x},\widetilde{t})$ as well as
the different frequencies contribute to the integral in (\ref{40}). The
contribution of the backward-moving wave, given by $\varphi(\widetilde
{x},\widetilde{x}_{i};\widetilde{\omega};-\widetilde{\omega}^{0})$, and
different frequencies depends on the value $\frac{v_{2}^{2}}{v_{1}^{2}%
}\widetilde{\sigma}_{x}^{2}/2$: the smaller this value the bigger contribution.

Let us consider the case of the incoming plane wave when $\widetilde{\sigma
}_{x}\rightarrow\infty$ (we remind that (\ref{40}) is valid for $\sigma
_{||}\rightarrow\infty$). Using again one-dimensional version of (\ref{33}),
we obtain
\begin{equation}
f_{\pm}(\widetilde{x},\widetilde{t})=-C\operatorname{Im}[e^{\mp i\widetilde
{\omega}^{0}\widetilde{t}}\varphi_{1}(\widetilde{x},;\widetilde{\omega}%
^{0}),\widetilde{\sigma}_{x}\rightarrow\infty, \label{43}%
\end{equation}
where $\varphi_{1}(\widetilde{x},\widetilde{x}_{i};\widetilde{\omega}^{0})$ is
defined by Eqs. (\ref{41}) with $\widetilde{\omega}=\widetilde{\omega}^{0}$and
multiplied by $v_{1}^{2}/v_{2}^{2}\widetilde{\sigma}_{x}$. Thus, in this case
there is no contribution of the backward-moving wave and frequencies different
from $\widetilde{\omega}^{0}$. Moreover, function (\ref{43}) ceases,
naturally, to depend on $\widetilde{x}_{i}$.

It is interesting to verify the result (\ref{43}) for the homogeneous case not
only in the parallel-to interface direction but also in the $x$-direction
($v_{1}=v_{2}=v_{3}$). We obtain%
\begin{equation}
f_{\pm}(x,t)=\frac{C}{2}\cos(\frac{\omega^{0}}{v_{1}}x\mp\omega^{0}%
t),v_{1}=v_{2}=v_{3}. \label{44}%
\end{equation}
Equation (\ref{44}) agrees with the result which can be obtained from
(\ref{12}) by employing the one dimensional "free" propagator for any phase
velocity $v_{i}$ (see, e.g.,\cite{Economou})
\begin{align}
\widetilde{g}_{0}(x,x^{\prime};t;\mathbf{k}_{||}  &  =0)=-\frac{i}{\pi}%
\int\limits_{0}^{\infty}d\omega\sin(\omega t)\left[  G_{0}^{+}(x,x^{/}%
;\omega^{2};\mathbf{k}_{||}=0)-G_{0}^{-}(x,x^{/};\omega^{2};\mathbf{k}%
_{||}=0)\right] \nonumber\\
&  =-\frac{1}{2v_{i}}\theta(v_{i}|t|-|x-x^{\prime}|)\tau(t), \label{45}%
\end{align}
where $\theta(v_{i}|t|-|x-x^{\prime}|)$ is the Heaviside step function and%
\begin{align}
\tau(t)  &  =\theta(t)-\theta(-t)=1,t>0,\nonumber\\
\tau(t)  &  =-1,t<0. \label{46}%
\end{align}

\begin{figure}
[ptb]
\begin{center}
\includegraphics[
height=2.7596in,
width=4.2211in
]%
{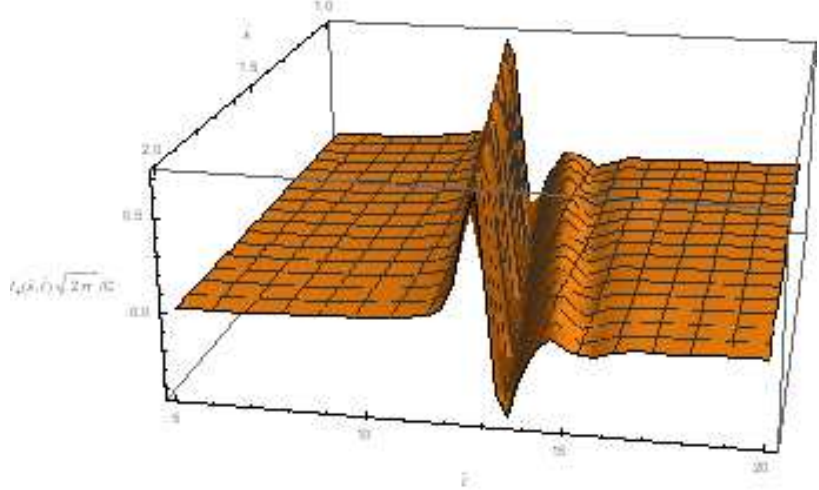}%
\caption{Wave packet with small dispersion transmitted through the spacer}%
\label{fig1}%
\end{center}
\end{figure}

\begin{figure}
[ptb]
\begin{center}
\includegraphics[
height=2.7968in,
width=3.8069in
]%
{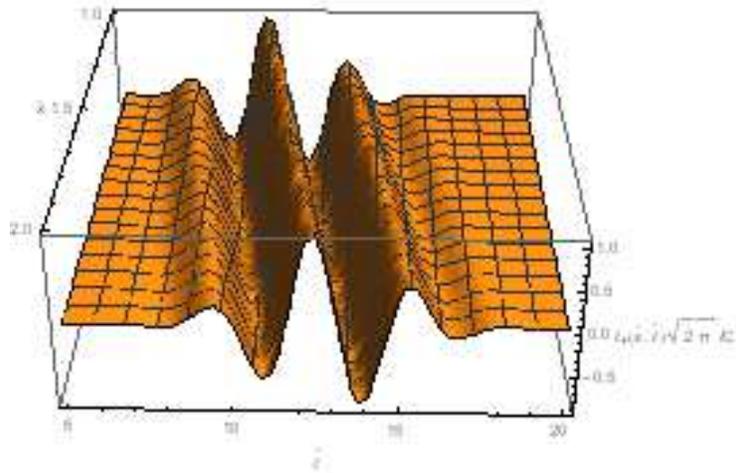}%
\caption{Same as in Figure \ref{fig1} but with bigger dispersion}%
\label{fig2}%
\end{center}
\end{figure}

We plotted the solution of (\ref{40}) for $f_{+}(\widetilde{x},\widetilde
{t})\sqrt{2\pi}/C$ in different spatial areas (in accordance with (\ref{41}))
for the case of symmetric threelayer ($v_{1}=v_{3}$) and different factor
$\frac{v_{2}^{2}}{v_{1}^{2}}\widetilde{\sigma}_{x}^{2}/2$ defining the
contribution of the backward-moving components of the incoming wave. Figure
\ref{fig1} shows the transmitted through the spacer wave (originated at
$\widetilde{x}_{i}=-5$) in the spatial interval $\widetilde{x}=1\div2$ (behind
the spacer) and as a function of time ($\widetilde{t}=5\div20$) at
$v_{1}=v_{3}=v_{2}/2$ and dispersion $\widetilde{\sigma}_{x}=0.2$
($\frac{v_{2}^{2}}{v_{1}^{2}}\widetilde{\sigma}_{x}^{2}/2=0.08$).
Dimensionless time of wave arrival from the starting point $\widetilde{x}%
_{i}=-5$ at the interface between the first and second layers can be estimated
as $\widetilde{t}_{a}=\left\vert x_{i}\right\vert /v_{1}t_{d}=\left\vert
\widetilde{x}_{i}\right\vert v_{2}/v_{1}=10$. The characteristic time
$t_{d}=d/v_{2}$ for the spacer of 100 nm thick is of the order of one
femtosecond ($10^{-7}m/10^{8}m\sec^{-1}=10^{-15}\sec$). Figure \ref{fig2}
presents the space-time dependence of the transmitted wave with the same
parameters except the bigger dispersion, i.e. at $\widetilde{\sigma}_{x}=1$
($\frac{v_{2}^{2}}{v_{1}^{2}}\widetilde{\sigma}_{x}^{2}/2=2$). We see from
Fig. \ref{fig1}, that when the parameter $\frac{v_{2}^{2}}{v_{1}^{2}%
}\widetilde{\sigma}_{x}^{2}/2$, related to the dispersion of the incoming wave
packet, is small, the backward-moving component of the wave packet and the
frequencies different from $\widetilde{\omega}^{0}$ result in a later arrival
of the wave packet to the points of interval $\widetilde{x}=1\div2$ and more
short in time transmission through the given point of the interval (as
compared to Fig. \ref{fig2}). Correspondingly, in the case when this parameter
$\frac{v_{2}^{2}}{v_{1}^{2}}\widetilde{\sigma}_{x}^{2}/2$ is bigger, the
contribution of the backward-moving wave and frequencies different from
$\widetilde{\omega}^{0}$ is small that leads to more earlier arrival to the
points of the selected spatial interval while the transmission through the
given spatial point becomes more extended in time and closer to the
transmission of the plane wave.

\begin{figure}
[ptb]
\begin{center}
\includegraphics[
height=3.4359in,
width=3.928in
]%
{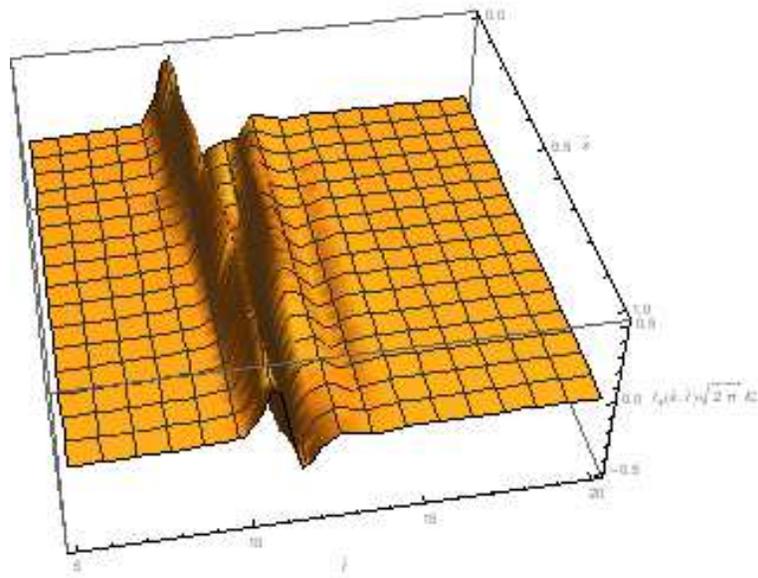}%
\caption{Wave packet propagation inside the spacer with dispersion as
in\ Figure \ref{fig1}}%
\label{fig3}%
\end{center}
\end{figure}

\begin{figure}
[ptb]
\begin{center}
\includegraphics[
height=2.9378in,
width=3.4497in
]%
{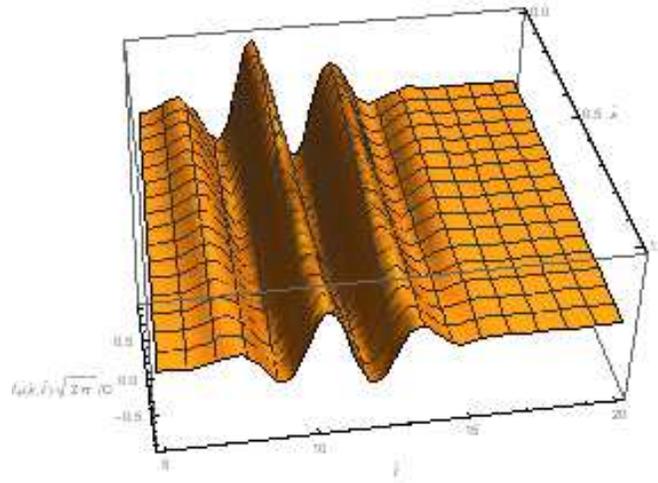}%
\caption{Wave packet propagation through the spacer with dispersion as in
Figure \ref{fig2}}%
\label{fig4}%
\end{center}
\end{figure}

It is of interest to have a look at the propagation in time of the wave packet
inside the spacer. Figures \ref{fig3}, \ref{fig4} show the propagation of the
wave packet from $\widetilde{x}=0$ to $\widetilde{x}=1$ (from one interface to
another) at the same parameters as in Figs. \ref{fig1}, \ref{fig2}. One can
see that the influence of the backward-moving component of the wave packet and
different frequencies is qualitatively the same as in Figs. \ref{fig1}%
,\ref{fig2} although in the case of small parameter $\frac{v_{2}^{2}}%
{v_{1}^{2}}\widetilde{\sigma}_{x}^{2}/2$ (Fig. \ref{fig3}), when this
influence is essential, the wave propagation shows more structured picture (as
compared to Fig. \ref{fig1}).

\begin{figure}
[ptb]
\begin{center}
\includegraphics[
height=2.7683in,
width=3.7983in
]%
{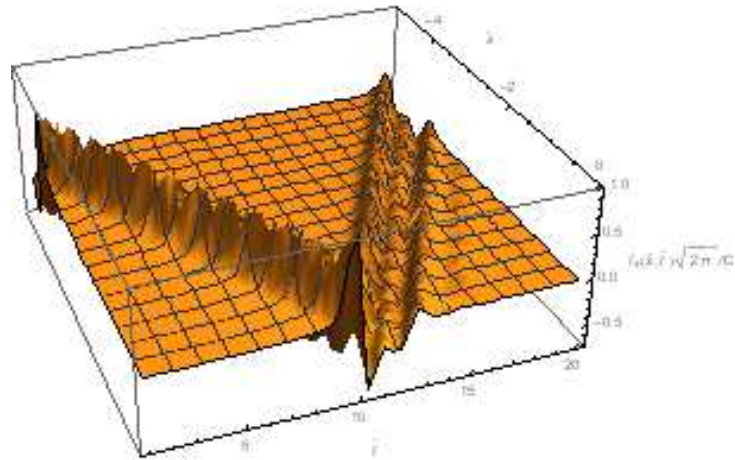}%
\caption{Space-time situation before the spacer for small wave packet
dispersion}%
\label{fig5}%
\end{center}
\end{figure}

\begin{figure}
[ptb]
\begin{center}
\includegraphics[
height=2.9421in,
width=3.5622in
]%
{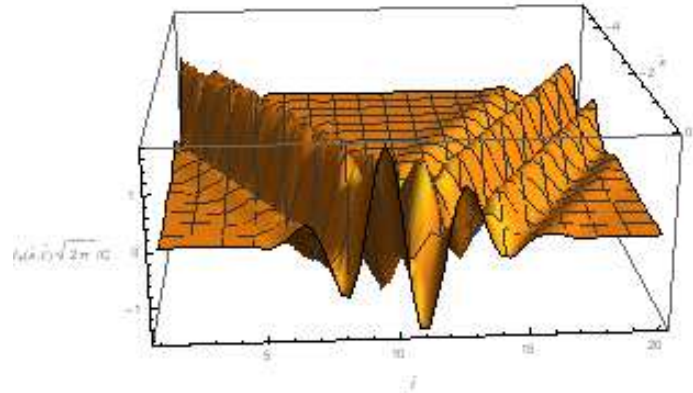}%
\caption{The same as in Figure \ref{fig5} but with increased dispersion}%
\label{fig6}%
\end{center}
\end{figure}

\begin{figure}
[ptb]
\begin{center}
\includegraphics[
height=2.943in,
width=3.2552in
]%
{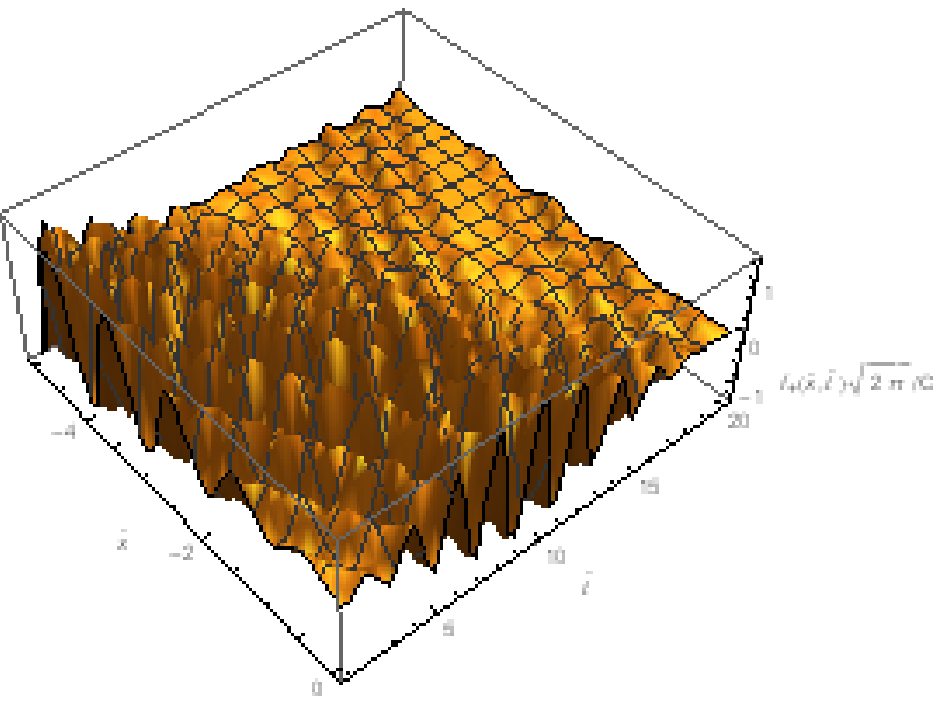}%
\caption{Space-time picture before the spacer when the contribution of the
reflected wave is small}%
\label{fig7}%
\end{center}
\end{figure}

At last, Figs. \ref{fig5}, \ref{fig6}, \ref{fig7} visualize the situation
before a spacer ($\widetilde{x}=-5\div0$). There are propagating and
reflecting waves contributing to the resulting picture. In Fig. \ref{fig5} one
can see the incoming and reflecting waves at small patameter $\frac{v_{2}^{2}%
}{v_{1}^{2}}\widetilde{\sigma}_{x}^{2}/2=0.08$. When this parameter increases,
the contribution of the reflected wave is more pronounced (Fig. \ref{fig6}).
However, as it follows from the last line of Eq. (\ref{41}), the reflection
amplitude $\widetilde{r}(\widetilde{\omega})=0$ when $\widetilde{\omega}=\pi
n$ ($n$ is integer and $v_{1}=v_{3}$). Thus, if the main contribution to the
integral (\ref{40}) comes from $\widetilde{\omega}\thickapprox$ $\widetilde
{\omega}^{0}=\pi n$, there is no reflecting wave. This situation is realized
closely for the large factor $\frac{v_{2}^{2}}{v_{1}^{2}}\widetilde{\sigma
}_{x}^{2}/2$ (see Eq. (\ref{43})) and is visualized for$\ \frac{v_{2}^{2}%
}{v_{1}^{2}}\widetilde{\sigma}_{x}^{2}/2=8$ (large dispersion $\widetilde
{\sigma}_{x}=2$) in Fig. \ref{fig7}.%

%TCIMACRO{\FRAME{ftbpFU}{4.2211in}{2.7596in}{0pt}{\Qcb{Wave packet with small
%dispersion transmitted through the spacer}}{\Qlb{fig1}}{fig1.eps}%
%width 4.2211in;  height 2.7596in;  depth 0pt;  original-width 4.7625in;
%original-height 3.1021in;  cropleft "0";  croptop "1";  cropright "1";
%cropbottom "0";  filename 'fig1.eps';file-properties "XNPEU";}} }%
%BeginExpansion
%EndExpansion
%

%TCIMACRO{\FRAME{ftbpFU}{3.8069in}{2.7968in}{0pt}{\Qcb{Same as in Figure
%\ref{fig1} but with bigger dispersion}}{\Qlb{fig2}}{fig2.eps}%
%width 3.8069in;  height 2.7968in;  depth 0pt;  original-width 3.7593in;
%original-height 2.7544in;  cropleft "0";  croptop "1";  cropright "1";
%cropbottom "0";  filename 'fig2.eps';file-properties "XNPEU";}} }%
%BeginExpansion
%EndExpansion
%

%TCIMACRO{\FRAME{ftbpFU}{3.928in}{3.4359in}{0pt}{\Qcb{Wave packet propagation
%inside the spacer with dispersion as in\ Figure \ref{fig1}}}{\Qlb{fig3}%
%}{fig3.eps}{\special{ language "Scientific Word";  type "GRAPHIC";
%width 3.928in;  height 3.4359in;  depth 0pt;  original-width 4.4287in;
%original-height 3.87in;  cropleft "0";  croptop "1";  cropright "1";
%cropbottom "0";  filename 'fig3.eps';file-properties "XNPEU";}} }%
%BeginExpansion
%EndExpansion
%

%TCIMACRO{\FRAME{ftbpFU}{3.4497in}{2.9378in}{0pt}{\Qcb{Wave packet propagation
%through the spacer with dispersion as in Figure \ref{fig2}}}{\Qlb{fig4}%
%}{fig4.eps}{\special{ language "Scientific Word";  type "GRAPHIC";
%width 3.4497in;  height 2.9378in;  depth 0pt;  original-width 4.0248in;
%original-height 3.4238in;  cropleft "0";  croptop "1";  cropright "1";
%cropbottom "0";  filename 'fig4.eps';file-properties "XNPEU";}} }%
%BeginExpansion
%EndExpansion
%

%TCIMACRO{\FRAME{ftbpFU}{3.7983in}{2.7683in}{0pt}{\Qcb{Space-time situation
%before the spacer for small wave packet dispersion}}{\Qlb{fig5}}%
%{fig5.eps}{\special{ language "Scientific Word";  type "GRAPHIC";
%width 3.7983in;  height 2.7683in;  depth 0pt;  original-width 4.248in;
%original-height 3.0874in;  cropleft "0";  croptop "1";  cropright "1";
%cropbottom "0";  filename 'fig5.eps';file-properties "XNPEU";}} }%
%BeginExpansion
%EndExpansion
%

%TCIMACRO{\FRAME{ftbpFU}{3.5622in}{2.9421in}{0pt}{\Qcb{The same as in Figure
%\ref{fig5} but with increased dispersion}}{\Qlb{fig6}}{fig6.eps}%
%width 3.5622in;  height 2.9421in;  depth 0pt;  original-width 4.3033in;
%original-height 3.5483in;  cropleft "0";  croptop "1";  cropright "1";
%cropbottom "0";  filename 'fig6.eps';file-properties "XNPEU";}} }%
%BeginExpansion
%EndExpansion
%

%TCIMACRO{\FRAME{ftbpFU}{3.2552in}{2.943in}{0pt}{\Qcb{Space-time picture before
%the spacer when the contribution of the reflected wave is small}}{\Qlb{fig7}%
%}{fig7.eps}{\special{ language "Scientific Word";  type "GRAPHIC";
%width 3.2552in;  height 2.943in;  depth 0pt;  original-width 3.7593in;
%original-height 3.3961in;  cropleft "0";  croptop "1";  cropright "1";
%cropbottom "0";  filename 'fig7.eps';file-properties "XNPEU";}} }%
%BeginExpansion
%EndExpansion

\section{Conclusion}

We have considered the wave equation for a layered structure made of a layer
(spacer) sandwiched between two semi-infinite layers all with different
constant wave phase velocity changing while crossing the interfaces between
layers. The solution to the wave equation with the one-dimensional (in the
perpendicular-to-interfaces direction) dependency of the phase velocity is
considered by means of finding the space-time propagator for this second-order
in time equation. The propagator is obtained by calculating the Green function
for the wave equation using the multiple-scaterring theory approach.
Scatterings at the interfaces leading to propagation through and reflection
from interfaces are described as scatterings by the corresponding interface
localized effective potentials. The obtained propagator gives the exact
solution for the considered wave equation in the layered system for any
initial value. The specific solution for the initial Gaussian wave packet is
obtained and analyzed. The general solution shows the contribution of both the
forward- and backward-moving components of the initial wave packet. The
contribution of the backward-moving components and different frequencies to
the solution is defined by the factor related to the dispersion of the initial
wave packet. The obtained solution allows for numerical visualization of the
processes of transmission through and reflection from the threelayer in time,
which is performed for the simple case of the symmetric threelayer and
perpendicular-to-interfaces direction of the incoming wave packet.

\end{document}